\documentclass[12pt]{article}

\let\OLDthebibliography\thebibliography
\renewcommand\thebibliography[1]{
  \OLDthebibliography{#1}
  \setlength{\parskip}{0pt}
  \setlength{\itemsep}{6.1pt plus 0.3ex}
}

\usepackage[left=2.8cm,right=2.8cm,top=2.8cm,bottom=2.8cm]{geometry}

\usepackage[colorlinks=true,linkcolor=black,citecolor=black,urlcolor=blue,filecolor=black]{hyperref}

\usepackage{amsmath}
\usepackage{amsfonts}
\usepackage{amssymb}
\usepackage{stmaryrd}
\usepackage{mathtools}
\usepackage{mathrsfs}
\usepackage{color}
\usepackage{soul}
\usepackage[nosort]{cite}
\usepackage[normalem]{ulem}
 \csname
@addtoreset\endcsname{equation}{section}

\def\d{\delta}

\def\e{\epsilon}

\def\Th{\Theta}

\def\l{\lambda}
\def\L{\Lambda}
\def\m{\mu}
\def\n{\nu}

\def\r{\rho}

\def\vf{\varphi}

\def\o{\omega}
\def\O{\Omega}

\def\cC{{\cal C}}

\def\cH{{\cal H}}

\def\cL{{\cal L}}
\def\cM{{\cal M}}

\def\cO{{\cal O}}
\def\cP{{\cal P}}

\def\cS{{\cal S}}

\def\cU{{\cal U}}

\def\cW{{\cal W}}

\def\CC{\mathbb{C}}

\def\RR{\mathbb{R}}

\def\ZZ{\mathbb{Z}}

\def\be{\begin{equation}}
\def\ee{\end{equation}}
\def\bea{\begin{eqnarray}}
\def\eea{\end{eqnarray}}
\def\ba{\begin{array}}
\def\ea{\end{array}}

\def\nn{\nonumber}

\def\bra{\langle}
\def\ket{\rangle}

\begin{document}

\vspace{30pt}

\begin{center}


{\Large\sc BMS Modules in Three Dimensions}

---------------------------------------------------------------------------------------------------------


\vspace{25pt}
{\sc A.~Campoleoni${}^{\; a,}$\footnote{Postdoctoral Researcher of the Fund for Scientific Research-FNRS 
Belgium.}, H.A.~Gonzalez${}^{\; a}$, B.~Oblak${}^{\; a,b,}$\footnote{Research fellow of the Fund for 
Scientific Research-FNRS Belgium.} and M.~Riegler${}^{\; c}$}

\vspace{10pt}
{${}^a$\sl\small
Universit{\'e} Libre de Bruxelles\\
and International Solvay Institutes\\
ULB-Campus Plaine CP231\\
1050 Brussels,\ Belgium
\vspace{10pt}

${}^b$\sl\small 
DAMTP, Centre for Mathematical Sciences\\
University of Cambridge\\
Wilberforce Road\\
Cambridge CB3 0WA,\ United Kingdom
\vspace{10pt}

${}^c$\sl\small 
Institute for Theoretical Physics\\
Vienna University of Technology\\
Wiedner Hauptstrasse 8-10\\
1040 Vienna,\ Austria
\vspace{10pt}

{\it andrea.campoleoni@ulb.ac.be, hgonzale@ulb.ac.be,\\ boblak@ulb.ac.be, rieglerm@hep.itp.tuwien.ac.at} 
}

\vspace{50pt} {\sc\large Abstract} \end{center}

\noindent
We build unitary representations of the BMS algebra and its 
higher-spin extensions in three dimensions, using induced representations as a guide. Our prescription 
naturally emerges from an ultrarelativistic limit of highest-weight representations of  Virasoro and $\cW$ 
algebras, which is to be contrasted with non-relativistic limits that typically give non-unitary 
representations. To support this dichotomy, we also point out that the ultrarelativistic and non-relativistic 
limits of generic $\cW$ algebras differ in the structure of their non-linear terms.


\newpage


\tableofcontents


\section{Introduction}\label{sec:intro}

It has long been known that asymptotically flat gravitational theories in three and four space-time 
dimensions enjoy powerful symmetries at null infinity, given by an infinite-dimensional extension of the 
Poincar\'e group known as the Bondi-Metzner-Sachs (BMS) group 
\cite{Bondi:1962px,Sachs:1962wk,Ashtekar:1996cd,Barnich:2006av}. The latter and its local generalisation 
\cite{Barnich:2009se} have been the focus of renewed interest in the last few years due to their relation 
e.g.\ to holography \cite{Barnich:2010eb}, soft graviton theorems \cite{Strominger:2013jfa} and black holes 
\cite{Hawking:2016msc}.\\

In three space-time dimensions this extension of Minkowski isometries is closely related to the 
infinite-dimensional symmetry enhancement of Anti-de Sitter space at spatial infinity \cite{Brown:1986nw}.  In 
the latter case asymptotic symmetries are generated by two copies of the Virasoro algebra and admit an 
\.In\"on\"u-Wigner contraction that reproduces the $\mathfrak{bms}_3$ algebra.\\

In three dimensions one can also broaden the purely gravitational setup to include ``higher-spin'' gauge 
fields\footnote{The little group of massless particles does not admit arbitrary discrete helicities in $D=3$, 
but in this context it is customary to use the word ``spin'' to label the representations of the Lorentz group 
under which fields transform.} on both flat and AdS backgrounds. The equations that, for $D > 3$, describe the 
propagation of a massless field of spin $s$ indeed imply the absence of local degrees of freedom in $D = 3$ 
when $s \geq 2$. This feature is manifest in the Chern-Simons formulation of both Einstein gravity 
\cite{Achucarro:1987vz,Witten:1988hc} and higher-spin theories \cite{Blencowe:1988gj}. In this approach one 
can also accommodate flat backgrounds \cite{Campoleoni:2011tn,Afshar:2013vka,Gonzalez:2013oaa}, thus bypassing 
the no-go results severely constraining higher-spin interactions in $D>3$ Minkowski space (see e.g.\ 
\cite{Bekaert:2010hw}). The presence of higher-spin fields further enhances the asymptotic symmetries. Around 
AdS these generically consist of two copies of a non-linear $\cW$ algebra 
\cite{Henneaux:2010xg,Campoleoni:2010zq}; around flat space they consist of a non-linear higher-spin extension 
of the $\mathfrak{bms}_3$ algebra \cite{Afshar:2013vka,Gonzalez:2013oaa}, which can be obtained as a 
contraction of the direct sum of two $\cW$ algebras.\\

Identifying a symmetry does not suffice to control its implementation at the quantum level: to this end one 
also needs to select the representations that are physically relevant in a given context.
With this motivation in mind, in this note we study a class of representations of the $\mathfrak{bms}_3$ 
algebra and of its higher-spin extensions.
In the gravitational case, the representations we are going to describe are the algebraic counterpart of the 
unitary representations of the BMS$_3$ group built and classified in \cite{Barnich:2014kra,Barnich:2015uva} 
along the lines originally used by Wigner for the Poincar\'e group \cite{Wigner:1939}. The Hilbert space of 
each such representation consists of wavefunctionals in supermomentum space, in direct analogy to standard 
quantum mechanics. In the higher-spin case, analogous induced representations were built in 
\cite{Campoleoni:2015qrh} and their Lie-algebraic version was briefly investigated. The main purpose of the 
present paper is to delve deeper in the details of that algebraic construction.\\

We stress that the inclusion of higher-spin fields is a highly non-trivial extension of the basic setup, due 
to the non-linearities that appear in the symmetry algebra on both AdS and flat backgrounds. As a result, 
standard group-theoretic methods fail to apply. One faces a similar situation when analyzing extended BMS 
symmetry in four dimensions: representations of the globally well defined BMS group have been classified 
\cite{McCarthy517,McCarthy317}, but a corresponding result for its local counterpart is still missing. Aside 
from their applications in three-dimensional higher-spin theories, we thus hope that our techniques will also 
prove useful in this challenging domain.\\

We will first build representations in a basis suggested by the theory of induced representations of Lie 
algebras, before showing how one can move to a basis of supermomentum eigenstates analogous to one-particle 
states with definite momentum. For that reason, in sect.~\ref{sec:poincare} we briefly review the construction 
of representations of the three-dimensional Poincar\'e algebra in a way that simplifies generalisations to the 
$\mathfrak{bms}_3$ algebra and its higher-spin extensions. We also recall how standard unitary representations 
of the Poincar\'e algebra emerge from an ultrarelativistic limit of highest-weight representations of 
$\mathfrak{so}(2,2)$. We then apply a similar construction to the $\mathfrak{bms}_3$ algebra in 
sect.~\ref{sec:bms}, and to its higher-spin extensions in sect.~\ref{sec:hs}. In both cases we also comment on 
the emergence of unitary representations from an ultrarelativistic limit of highest-weight representations of 
the (extended) local conformal algebra, while stressing that non-relativistic limits naturally lead to 
non-unitary representations as those considered in \cite{Bagchi:2009pe,Grumiller:2014lna}.

\section{Poincar\'e modules in three dimensions}\label{sec:poincare}

The unitary representations of the Poincar\'e group in three dimensions were classified in 
\cite{Binegar:1981gv} and recently reviewed e.g.\ in \cite{Barnich:2014kra} due to their relation with BMS$_3$ 
representations. Here we discuss how Wigner's standard method for the construction of irreducible, unitary 
representations of the Poincar\'e group (as presented e.g.\ in \cite{weinberg1995quantum}) can be recovered 
from induced representations of the Poincar\'e \emph{algebra}, also known as Poincar\'e modules. The advantage 
of this approach is that such modules can also be built for the $\mathfrak{bms}_3$ algebra and its non-linear 
higher-spin generalisations.

\subsection{The Poincar\'e algebra}

In three dimensions, the Lie algebra of the Poincar\'e group is spanned by three Lorentz generators 
$J_m$ and three translation generators $P_m$ ($m=-1,0,1$) whose Lie brackets read
\begin{subequations}\label{pal}
\bea
\,[J_m,J_n] & = & (m-n)\,J_{m+n}\,,\\
\,[J_m,P_n] & = & (m-n)\,P_{m+n}\,,\\
\,[P_m,P_n] & = & 0\,.
\eea
\end{subequations}
Our conventions are such that these basis elements generate the complexification of the Poincar\'e algebra. Real 
translations, for example, are generated by linear combinations $\alpha_m P_m$ with complex coefficients satisfying 
$(\alpha_m)^*=\alpha_{-m}$; similarly real boosts are generated by combinations $zJ_1+z^*J_{-1}$ while 
rotations are generated by $\theta J_0$, with $\theta$ real. Accordingly, in any unitary representation, the 
operators representing Poincar\'e generators must satisfy the hermiticity conditions
\be \label{herm}
(P_m)^{\dagger}=P_{-m}\,,
\quad
(J_m)^{\dagger}=J_{-m}\,.
\ee
Note that the $P_m$'s correspond to the standard translation generators $\cP^{\mu}$ (with $\mu=0,1,2$ a 
Lorentz index) as $P_0=\cP^0$, $P_1=\cP^1+i\cP^2$ and $P_{-1}=\cP^1-i\cP^2$.\\ 

The three-dimensional Poincar\'e algebra is thus the semi-direct sum
\be
\mathfrak{iso}(2,1)
=
\mathfrak{sl}(2,\mathbb{R})
\inplus_{\text{ad}}\left(\mathfrak{sl}(2,\mathbb{R})\right)_{\text{Ab}}
\label{palge}
\ee
where $\mathfrak{sl}(2,\mathbb{R})\cong\mathfrak{so}(2,1)$ is the Lorentz algebra (generated by $J_m$'s) and 
$\left(\mathfrak{sl}(2,\mathbb{R})\right)_{\text{Ab}}$ is an Abelian Lie algebra of translations (generated by 
$P_m$'s) isomorphic to the Lorentz algebra as a vector space, and acted upon by Lorentz transformations 
according to the adjoint representation. The Poincar\'e algebra admits two quadratic Casimir operators: the 
mass squared
\be
\cM^2=P_0^2-P_1P_{-1}
\label{msq}
\ee
and the three-dimensional analogue of the square of the Pauli-Lubanski vector,
\be
\cS
=
P_0J_0-\frac{1}{4}\left(J_1P_{-1}+J_{-1}P_1+P_1J_{-1}+P_{-1}J_1\right).
\label{casij}
\ee
The eigenvalues of these operators can be used to classify irreducible representations.

\subsection{Poincar\'e modules}\label{sec:p-modules}

Irreducible unitary representations of the Poincar\'e group are obtained by considering the orbit of a given 
momentum under Lorentz transformations --- i.e.\ all momenta $p^{\mu}=(p^0,p^1,p^2)$ that satisfy $p^2 = -M^2$ 
for some mass $M$ --- and building a Hilbert space of wavefunctions on that orbit. The eigenvalue of $P_0$ 
gives the energy of the corresponding particle and inspection of \eqref{pal} shows that the operators that 
commute with $P_0$ are $P_{1}$, $P_{-1}$ and $J_0$. It is therefore natural to build a basis of eigenstates 
of momentum for the Hilbert space of wavefunctions on the orbit; we will denote such eigenstates by $|p^\mu, 
s\ket$. These correspond to plane waves of definite momentum $p^{\mu}$, while $s\in\RR$ is a spin label 
related to the eigenvalue of $J_0$ in a particular frame (see eq.~\eqref{rest1}). Different values of $s$ 
yield inequivalent irreducible representations  \cite{Binegar:1981gv,Barnich:2014kra}. Under a Lorentz 
transformation parametrised by $\L^{\m}{}_{\n}$ these wavefunctions transform as 
\be \label{trans}
 U(\Lambda)|p^\mu,s\rangle = e^{is\theta}|\Lambda^\mu{}_{\nu}p^\nu,s\rangle \,,
\ee
where $U(\Lambda)$ is a unitary operator and $\theta$ is a $\Lambda$-dependent Wigner angle. The components 
$p^{\mu}$ with $\mu=0,1,2$ are related to the eigenvalues $p_m$ of the generators $P_m$ by $p^0=p_0$, 
$p^1=(p_1+p_{-1})/2$ and $p^2=(p_1-p_{-1})/2i$.\\

Lorentz transformations act transitively on the momentum orbit, so for each fixed value of the mass squared 
one can choose a ``standard'' momentum $k^\mu$ and obtain all plane waves by acting with Lorentz boosts on the 
corresponding wavefunction $|k^\mu,s\ket$. For \emph{massive} representations --- on which we focus for the 
sake of comparison with $\mathfrak{bms}_3$ and its higher-spin extensions --- one can choose as a 
representative the momentum $k^{\mu}=(M,0,0)$ of the particle at rest. We denote by $|M,s\ket$ the 
corresponding wavefunction, which satisfies
\be\label{rest1}
P_0 |M,s\ket = M |M,s\ket \, , \quad 
P_{-1} |M,s\ket = P_1 |M,s\ket = 0 \, , \quad 
J_0 |M,s\ket = s |M,s\ket \,,
\ee
and call it the \emph{rest-frame state} of the representation. To obtain a plane wave $|p^{\mu},s\rangle$ 
with boosted momentum, one can act with a Lorentz transformation implemented by the unitary operator
\be
U(\omega)
=
\exp\left[\,i\left(\omega J_1+\omega^*J_{-1}\right)\right] ,
\label{lobast}
\ee
where $\o$ is the complex rapidity
\be \label{rapidity}
\omega
=
\frac{i}{2}\,\text{arcsinh}
\left(
\frac{\sqrt{p_1p_{-1}}}{M}
\right)
\frac{p_{-1}}{\sqrt{p_1p_{-1}}} 
\ee
that one can obtain by inverting the relation $p^\mu = \L^\mu{}_\nu k^\nu$ taking into account \eqref{herm}.\\

The previous discussion is standard, but note that \eqref{rest1} defines a one-dimensional representation of 
the subalgebra generated by $\{P_m,J_0\}$. Given a representation of a subalgebra $\mathfrak{h}$ of the Lie 
algebra $\mathfrak{g}$ on a vector space $V$, one can always build a representation of $\mathfrak{g}$ on a 
suitable quotient of the space $\cU(\mathfrak{g}) \otimes V$, where $\cU(\mathfrak{g})$ is the universal 
enveloping algebra of $\mathfrak{g}$ (see e.g.\ sect.~10.7 of \cite{schottenloher2008mathematical}).  
Representations of this kind are called \emph{induced modules}. With this method one can construct an 
irreducible representation of the Poincar\'e \emph{algebra} on the vector space $\cH_{M}$ with basis 
vectors\footnote{The states \eqref{boost1} form a basis of the induced $\mathfrak{iso}(2,1)$-module
\[
\textrm{Ind}^{\mathfrak{iso}(2,1)}_\mathfrak{h}(\rho)
\equiv
\left( \cU(\mathfrak{\mathfrak{p}_3}) \otimes \CC \right) / 
\left\{ X \otimes \l - 1 \otimes \r[X] \l \,|\, X \in \mathfrak{h} ,\, \l \in \CC \right\} ,
\]
where $\mathfrak{h} = \textrm{Span}\{P_n,J_0\}$ and $\r$ is the one-dimensional $\CC$-valued representation
\[
\r[P_0] = M \, , \quad 
\r[P_{-1}] = \r[P_1] = 0 \, , \quad 
\r[J_0] = s 
\]
defined by \eqref{rest1}. The quotient amounts to the rule that when one acts by the left with any element in 
$\mathfrak{iso}(2,1)$ one moves $P_m$ and $J_0$ to the right by computing commutators and then lets them act 
on $|M,s\ket$, as is implicit in the ket notation \eqref{boost1}. \label{induced-foot}}
\be \label{boost1}
|k,l\,\ket = (J_{-1})^k (J_{1})^l |M,s\ket \, .
\ee
Upon acting from the left on the states \eqref{boost1} one obtains indeed linear operators on $\cH_M$ whose 
commutators coincide with \eqref{pal}.
Moreover, the Casimir operators \eqref{msq} and \eqref{casij} have the same eigenvalue on each state 
\eqref{boost1}, since they commute by construction with all elements of the algebra. This readily implies that 
the representation thus obtained is irreducible.\\

Unitarity, on the other hand, is far less obvious: it is not clear how to define a scalar product on the 
space $\cH_M$ spanned by the states \eqref{boost1}, even after enforcing the standard hermiticity conditions 
\eqref{herm}. Fortunately, experience with the Poincar\'e group suggests a way to circumvent the 
problem.\footnote{In sect.~\ref{Repsl2} we will also see an alternative way to define a scalar product on 
Poincar\'e modules from limits of representations of the $\mathfrak{so}(2,2)$ algebra.} Upon acting on the 
rest frame state $|M,s\ket$ with a Lorentz boost \eqref{lobast} one obtains (possibly up to an irrelevant 
phase) a plane wave 
\be \label{trans2}
|p^\mu,s\ket = U(\Lambda) |M,s\ket \,.
\ee
Here $p^\mu$ can be any momentum belonging to the orbit with mass $M$, provided one chooses properly the 
Lorentz parameter $\L$ as in \eqref{rapidity}. Such plane wave states can be normalised so that
\be \label{norm-waves}
\bra\, p^\mu, s \,|\, q^\mu , s \,\ket = \d_\mu(p,q) \, ,
\ee
where $\d_\mu$ is the Dirac distribution associated with the Lorentz-invariant measure
\be
d\m(q) =
\frac{dq_1 dq_{-1}}
{2i\sqrt{M^2+q_1q_{-1}}}\,.
\ee

In mapping the rest-frame state onto the states $|p^\mu,s\ket$ we applied \emph{finite} Lorentz 
transformations, so that we secretly brought the discussion back to the group-theoretic level. Nevertheless, 
to perform the ``change of basis'' from states of the form \eqref{boost1} to eigenstates of momentum, one does 
not need to control the full group structure; rather, it suffices to ensure that the boost \eqref{lobast} is 
well defined and that one can define a measure on the momentum orbit such that \eqref{norm-waves} is satisfied 
(see e.g.\ \cite{Barnich:2014kra} for more details). The states obtained by acting with boosts on $|M,s\ket$ 
can then be seen as infinite linear combinations of states \eqref{boost1}. Unitarity finally follows from the 
fact that plane waves form an orthonormal basis of the Hilbert space (cf.\ eq.~\eqref{norm-waves}).

\subsection{Ultrarelativistic limit of $\mathfrak{sl}(2,\mathbb{R})$ modules} \label{Repsl2}

In addition to being convenient for generalisations to infinite-dimensional extensions of the Poincar\'e 
algebra, Poincar\'e modules can be seen to arise as a limit of  unitary representations of the AdS$_3$ 
isometry algebra, namely $\mathfrak{so}(2,2)$. Owing to the isomorphism $\mathfrak{so}(2,2) \cong 
\mathfrak{sl}(2,\mathbb{R}) \oplus \mathfrak{sl}(2,\mathbb{R})$, the generators of this algebra can be divided 
in two groups, $\cL_m$ and $\bar{\cL}_m$ with \mbox{$m=-1,0,1$}, and their Lie brackets read
\be \label{so(2,2)}
[\cL_m,\cL_n]= (m-n)\, \cL_{m+n} \, , \qquad [\bar{\cL}_m,\bar{\cL}_n]= (m-n)\, \bar{\cL}_{m+n} \, .
\ee
As in (\ref{pal}) our conventions are such that this is a basis of the complexification of 
$\mathfrak{sl}(2,\mathbb{R})$, so that real 
$\mathfrak{sl}(2,\mathbb{R})$ matrices are linear combinations $i\,x_m\cL_m$ with $(x_m)^*=x_{-m}$. In 
particular, in any unitary representation the operators representing the generators $\cL_m$ and $\bar\cL_m$ 
must satisfy the hermiticity conditions
\be
(\cL_m)^{\dagger}=\cL_{-m}\,,
\qquad
(\bar{\cL}_m)^{\dagger}=\bar{\cL}_{-m}\,.
\label{kermit}
\ee
In terms 
of these basis elements the quadratic Casimir of each copy of $\mathfrak{sl}(2,\mathbb{R})$ reads
\be
\cC
=
\cL_0^2
-\frac{1}{2}
\left(\cL_1\cL_{-1}+\cL_{-1}\cL_1\right).
\label{slc}
\ee

The Poincar\'e algebra \eqref{pal} can be recovered from an \.In\"on\"u-Wigner contraction of \eqref{so(2,2)} 
by introducing a lenght scale $\ell$ (to be identified with the AdS radius) and by defining the new generators
\be
P_m\equiv\frac{1}{\ell}\left(\cL_m+\bar\cL_{-m}\right),
\qquad
J_m\equiv\cL_m-\bar\cL_{-m} \, .
\label{lpj}
\ee
The Lie brackets of $\mathfrak{sl}(2,\mathbb{R}) \oplus \mathfrak{sl}(2,\mathbb{R})$ are turned into
\begin{subequations}
\bea
\,[J_m,J_n] & = & (m-n)\,J_{m+n} \,, \\
\label{sll}
\,[J_m,P_n] & = & (m-n)\,P_{m+n} \,,\\
\,[P_m,P_n] & = & \ell^{-2}(m-n)\,J_{m+n} \,, \label{sll-3}
\eea
\end{subequations}
and in the limit $\ell \to \infty$ one recovers the Poincar\'e algebra. In addition 
the quadratic Casimir (\ref{slc}) can be combined with its counterpart 
$\bar{\cal C}$ in the second copy of $\mathfrak{sl}(2,\mathbb{R})$, producing
\be
\frac{2}{\ell^2} \left( {\cal C}+\bar{\cal C} \right)
= \cM^2 + \cO(\ell^{-2}) \, ,
\qquad
\frac{1}{\ell} \left({\cal C}-\bar{\cal C}\right)
= {\cal S} \, ,
\label{casilim}
\ee
where $\cM^2$ and $\cS$ are the Poincar\'e Casimirs \eqref{msq} and \eqref{casij}.\\

Aside from comparing Casimir operators, one can track how Poincar\'e modules (defined by \eqref{rest1} and 
\eqref{boost1}) emerge from the corresponding limit of highest-weight representations of $\mathfrak{so}(2,2)$. 
These are built out of highest-weight representations of $\mathfrak{sl}(2,\mathbb{R})$, which are defined 
starting from a  
state $|h\rangle$ that satisfies the conditions
\be
\cL_0|h\rangle=h\,|h\rangle\,,
\qquad
\cL_1|h\rangle=0\,.
\label{hwc}
\ee
The carrier space of the representation is then spanned by all descendant states 
$(\cL_{-1})^m|h\rangle$,\footnote{Note that  $\mathfrak{sl}(2,\RR)$ highest-weight representations can also be 
interpreted as induced modules. Eq.~\eqref{hwc} defines indeed a one-dimensional representation of the 
subalgebra spanned by $\{\cL_0,\cL_1\}$, while the vector space of descendant states can be identified with a 
quotient of $\cU(\mathfrak{sl}(2,\RR)) \otimes \CC$ as discussed in footnote \ref{induced-foot}. The main 
difference with respect to the Poincar\'e case is the splitting of the algebra as $\mathfrak{n^-} \oplus 
\mathfrak{h} \oplus \mathfrak{n}^+$, where $\mathfrak{n}^\pm$ are nilpotent subalgebras. This decomposition 
allows one to define a scalar product by enforcing the hermiticity condition \eqref{kermit}. One can then 
verify that $\bra h | (\cL_{1})^m  (\cL_{-1})^m | h \ket$ is positive for $h > 0$.} and the Casimir 
(\ref{slc}) takes the value $h(h-1)$. If one builds a similar representation
with weight $\bar h$ for a second copy of 
$\mathfrak{sl}(2,\mathbb{R})$, one can produce a 
representation of $\mathfrak{sl}(2,\mathbb{R})\oplus\mathfrak{sl}(2,\mathbb{R})$ from the tensor product.\\

To relate this tensor product --- spanned by the states $(\cL_{-1})^m (\bar{\cL}_{-1})^n|h,\bar{h}\ket$ --- 
to a Poincar\'e module, we rewrite it in the new basis given by \eqref{boost1}, where $M$ and $s$ are related 
to the $\mathfrak{so}(2,2)$ weights as
\be
M\equiv\frac{h+\bar h}{\ell}\,,
\qquad
s\equiv h-\bar h \, ,
\label{ms}
\ee
since in terms of the operators \eqref{lpj} one has
\be
P_0|h, \bar h\rangle
=
\frac{h+\bar h}{\ell}|h, \bar h\rangle\,, \qquad
J_0|h,\bar h\rangle
=
(h-\bar h)|h, \bar h\rangle\,.
\ee
This change of basis is invertible because no $J_n$ annihilate the vacuum. Each $\mathfrak{so}(2,2)$ 
representation now takes the form
\begin{subequations} \label{matrix-elements}
\bea
P_n |k,l\ket & = & \sum_{k',l'} \mathsf{P}^{(n)}_{k',l';\,k,l}(M,s,\ell) |k',l'\ket \,,\label{matrix-p} \\
J_n |k,l\ket & = & \sum_{k',l'} \mathsf{J}^{(n)}_{k',l';\,k,l}(M,s) |k',l'\ket
\eea
\end{subequations}
where $\mathsf{P}^{(n)}$ and $\mathsf{J}^{(n)}$ are infinite matrices and where only negative powers of 
$\ell$ appear in \eqref{matrix-p}. These only arise because of the highest-weight conditions, which can be 
rewritten as
\be \label{hwcPoincare}
\left(
P_{\pm1}\pm\frac{1}{\ell}J_{\pm1}
\right)|h, \bar h\rangle
= 0 \, ,
\ee
allowing to express the action of $P_n$ in terms of the states $|k,l\ket$.
As a result, the matrix elements $\mathsf{P}^{(n)}_{k',l';\,k,l}$ and $\mathsf{J}^{(n)}_{k',l';\,k,l}$ have a 
well defined limit for $\ell \to \infty$. By computing the action of the generators $P_n$ and $J_n$ on the 
Poincar\'e module spanned by \eqref{boost1}, one obtains the same outcome provided that the conformal weights 
scale as
\be \label{scaleh}
h= \frac{M\ell+s}{2}+\l+{\cal O}(\ell^{-1}),
\quad
\bar h= \frac{M\ell-s}{2}+\l+{\cal O}(\ell^{-1}) \, ,
\ee 
where $\l$ is an arbitrary parameter independent of $\ell$.
In particular, the latter condition implies that in the limit the $\mathfrak{sl}(2,\RR)$ highest-weight 
conditions \eqref{hwcPoincare} turn into the rest-frame conditions \eqref{rest1}. Note also that the 
Poincar\'e Casimirs $\cM^2$ and ${\cal S}$ take the values dictated by the rest frame conditions. Moreover, in 
principle one could also define a scalar product on Poincar\'e modules starting from the limit of the scalar 
product $\bra h | (\cL_{1})^m  (\cL_{-1})^n | h \ket$. This procedure will lead in general to a complicated 
non-diagonal quadratic form. We already know, however, that the plane-wave basis \eqref{norm-waves} 
diagonalises it, thus appearing as a natural alternative also from this vantage point.\\

Relation \eqref{scaleh} shows that the flat limit defined via \eqref{lpj} can be interpreted as an 
ultrarelativistic/high-energy limit from the viewpoint of AdS$_3$. 
Poincar\'e modules are thus remnants of $\mathfrak{so}(2,2)$ representations whose energy becomes large in 
the limit $\ell \to \infty$. In sect.~\ref{sec:galileo} we shall also discuss a different contraction from 
$\mathfrak{so}(2,2)$ to $\mathfrak{iso}(2,1)$, to be interpreted as a non-relativistic limit giving rise to 
representations of the type discussed in \cite{Bagchi:2009pe,Grumiller:2014lna}.

\section{Induced modules for the $\mathfrak{bms}_3$ algebra}\label{sec:bms}

In this section, we remark that one can readily obtain representations of the $\mathfrak{bms}_3$ algebra by 
exploiting the induced module construction introduced in sect.~\ref{sec:p-modules}. We then show how one can 
move to a basis of supermomentum eigenstates by following analogous steps to those that we reviewed for the 
Poincar\'e case. This basis then allows one to discuss the irreducibility and unitarity of the induced 
representations. We finally display how the previous representations can be obtained from an ultrarelativistic 
limit of Virasoro Verma modules, while recalling why Galilean limits typically lead to non-unitary 
representations of a different kind. 

\subsection{$\mathfrak{bms}_3$ algebra}

The $\mathfrak{bms}_3$ algebra is an infinite-dimensional algebra spanned by superrotation generators $J_m$ 
and supermomentum generators $P_m$ ($m \in \ZZ$) whose Lie brackets read
\begin{subequations} \label{jpc}
\bea
\,[J_m,J_n] & = & (m-n)J_{m+n}+\frac{c_1}{12}\,m(m^2-1)\,\delta_{m+n,0} \,,\\
\,[J_m,P_n] & = & (m-n)P_{m+n}+\frac{c_2}{12}\,m(m^2-1)\,\delta_{m+n,0} \,, \label{jp}\\
\,[P_m,P_n] & = & 0 \,,
\eea
\end{subequations}
where $c_1$ and $c_2$ are central charges. The central charge $c_2$ plays a key role for representation 
theory and it is e.g.\ non-vanishing in three-dimensional gravity \cite{Barnich:2006av}, where it takes the 
value $c_2 = \frac{3}{G}$ with $G$ being Newton's constant. The Poincar\'e algebra \eqref{pal} is a subalgebra 
of $\mathfrak{bms}_3$. Similarly to \eqref{palge}, the $\mathfrak{bms}_3$ algebra is the semi-direct sum 
\be
\mathfrak{bms}_3
=
\mathfrak{vir}\inplus_{\text{ad}}(\mathfrak{vir})_{\text{Ab}}
\label{bmsdef}
\ee
where $\mathfrak{vir}$ denotes the Virasoro algebra. In contrast with Poincar\'e, the operators \eqref{msq} 
and \eqref{casij} no longer commute with all generators of the algebra.\\

To the best of our knowledge, the classification of $\mathfrak{bms}_3$ Casimir operators is unknown. However, 
it was shown in \cite{feigin1983} that the only Casimirs of the Virasoro algebra are functions of its central 
charges. If one assumes that all $\mathfrak{bms}_3$ Casimirs can be obtained as flat limits of Virasoro 
Casimirs (in the same way that the Poincar\'e Casimirs can be seen as limits, cf.~\eqref{casilim}), then there 
are no $\mathfrak{bms}_3$ Casimirs other than its central charges.

\subsection{$\mathfrak{bms}_3$ modules}\label{sec:bms3modules}

Irreducible unitary representations of the BMS$_3$ group are classified by orbits of supermomenta under the 
action of finite superrotations, that is, by coadjoint orbits of the Virasoro group \cite{Barnich:2014kra}. In 
analogy with the Poincar\'e example, each orbit consists of supermomenta obtained by acting with 
superrotations on a given supermomentum $p$. The latter is a function on the circle,
\be \label{supermomentum}
p(\vf) = \sum_{n\in\mathbb{Z}}p_ne^{in\varphi} \, ,
\ee
and can be interpreted, from the gravitational viewpoint, as the Bondi mass aspect associated with an 
asymptotically flat metric in three dimensions --- i.e.\ the energy density carried by the gravitational field 
at null infinity. It transforms as a quadratic density (or equivalently as a CFT stress tensor) under 
superrotations.
The corresponding representation is then obtained by assuming the existence of a (quasi-)\-invariant measure 
on the orbit and by building a Hilbert space of square-integrable wavefunctionals on that orbit 
\cite{Barnich:2014kra}. This Hilbert space admits a basis of eigenstates of the operators $P_m$, which 
generalise plane waves of definite momentum, and that we will denote as $|p(\vf),s\ket$. Here $s\in\RR$ is a 
spin label directly analogous to its Poincar\'e counterpart; BMS$_3$ representations with identical 
supermomentum orbits but different spins are mutually inequivalent.

\subsubsection*{Massive modules}

An important class of representations is provided by supermomentum orbits that contain a constant $p(\vf) = M 
- c_2/24$, where $M > 0$ is a mass parameter and $c_2$ is the central charge entering \eqref{jpc}. This class 
contains e.g.\ the vacuum representation $M = 0$, that accounts for all perturbative boundary excitations 
around the vacuum \cite{Barnich:2015uva}. The Hilbert space of any such representation contains a wavefunction 
$|M,s\ket$ that satisfies
\be \label{rest-bms}
P_0|M,s\ket=M|M,s\ket\,,
\quad
P_m|M,s\ket=0
\;\text{for }m\neq0\,,
\quad
J_0|M,s\ket=s|M,s\ket\,,
\ee
i.e.\ which is a supermomentum eigenstate for the constant eigenvalue $p(\vf) = M - c_2/24$. In analogy with 
\eqref{rest1}, we will call $|M,s\ket$ the \emph{rest-frame state} of the representation.\\

As in the Poincar\'e case, \eqref{rest-bms} defines a one-dimensional representation of the subalgebra of 
\eqref{jpc} spanned by $\{P_n,J_0,c_1,c_2\}$. This representation can be used to define an induced 
$\mathfrak{bms}_3$ module $\cH_M$ with basis vectors
\be \label{eq:BoostedStatesDef}
J_{n_1}J_{n_2} \cdots J_{n_N}|M,s\rangle \, ,
\ee
where the $n_i$'s are non-zero integers such that $n_1\geq n_2\geq...\geq n_N$. With this ordering, states 
\eqref{eq:BoostedStatesDef} with different combinations of $n_i$'s are linearly independent within the 
universal enveloping algebra of $\mathfrak{bms}_3$, and acting on them from the left with the generators of 
the algebra provides linear operators on $\cH_M$ whose commutators coincide with \eqref{jpc}.\\

It is again unclear, however, how to define from scratch a scalar product on the space spanned by 
\eqref{eq:BoostedStatesDef}. Without scalar product one cannot look for null states to identify reducible 
modules, and the operators \eqref{msq} and \eqref{casij} can no longer be used to check irreducibility. 
Analogously to Poincar\'e, one can nevertheless consider the scalar product inherited from the limit of 
representations of the conformal algebra or, more naturally, reach a basis of supermomentum eigenstates. These 
states can be built (up to an irrelevant phase) from the rest-frame wavefunctional as
\be \label{eq:Superwaves}
|p(\vf),s\ket = U(\o) |M,s\ket \, ,
\ee
where\footnote{The exponential map from the Virasoro algebra to the Virasoro group is {\it not} locally 
surjective (see e.g.\ \cite{guieu2007algebre}), so exponential operators such as (\ref{expu}) cannot map the 
rest-frame state $|M,s\rangle$ on {\it all} other plane waves in the supermomentum orbit. Nevertheless, 
superrotations do act transitively on the orbit: 
\eqref{eq:Superwaves} is always correct for some unitary operator $U$, although the latter cannot always be 
written as an exponential \eqref{expu}. For the sake of simplicity, we assume that this subtlety does not 
affect our arguments; some additional comments can be found in appendix \ref{sesusu}.\label{footex}}
\be
U(\o) = \exp\left( i \sum_{n\in\ZZ^*} \o_n J_n \right),
\quad
\text{with }\omega_n^*=\omega_{-n}
\label{expu}
\ee
is a unitary operator implementing a finite superrotation. The complex coefficients $\o_n$ are the Fourier 
modes of a vector field on the circle $\o(\vf) \partial_\vf$. One readily verifies that the semi-direct 
structure (\ref{bmsdef}) implies that the states \eqref{eq:Superwaves} are eigenstates of supermomentum. 
Indeed, the Baker-Campbell-Hausdorff formula yields
\be \label{aqui}
P_m |p(\vf),s\ket = U \left( U^{-1} P_m U \right) |M,s\ket = U\cdot \exp\left( i\,\textrm{ad}_\o\right)[P_m] |M,s\ket \, ,
\ee
where $\sum_n \o_n J_n$ acts on $P_m$ according to the adjoint representation,
\be
\textrm{ad}_\o[P_m] = \sum_{n\in\ZZ^*} [\, \o_n J_n , P_m \,] \, .
\ee
The bracket (\ref{jp}) then implies that $U^{-1}P_mU$ is a certain combination of products of $P_m$'s and 
central charges $c_2$. This combination acts multiplicatively on the rest-frame state thanks to conditions 
\eqref{rest-bms}, from which we conclude as announced that $U|M,s\rangle$ is an eigenstate of supermomentum.\\

In fact we can even be more precise and say something about the eigenvalue of $U|M,s\rangle$ under $P_m$. 
(The complete argument is presented in appendix \ref{sesusu}, while here we only discuss its salient aspects.)
By construction, supermomenta transform under superrotations according to the coadjoint representation of the 
Virasoro group, which coincides with the standard transformation law of CFT stress tensors under conformal 
transformations. Explicitly, a superrotation $f(\vf)$ (satisfying $f(\vf+2\pi)=f(\vf)+2\pi$) maps a 
supermomentum $p(\vf)$ on a new supermomentum $(f\cdot p)(\vf)$ given by
\be
\label{coadjoint}
(f\cdot p)(f(\vf))=
\frac{1}{(f'(\vf))^2}\,
\left[p(\vf) +\frac{c_2}{12}\{f;\vf\}\right] \, ,
\ee
where $\{f;\vf\}$ is the Schwarzian derivative of $f$ at $\vf$.
Accordingly we know that the supermomentum of the state $U(\omega)|M,s\rangle$ takes the form 
\eqref{coadjoint} with $p(\vf)$ replaced by $M-c_2/24$, in which case the eigenvalue of $U|M,s\rangle$ under 
$P_m$ is the $m^{\text{th}}$ Fourier mode of $f\cdot p$. The diffeomorphism $f$ is the exponential of the 
vector field $\omega(\vf)\partial_{\vf}$; in other words $f$ is given by the flow of 
$\omega(\vf)\partial_{\vf}$ (see eq.\ \eqref{expif} for the exact correspondence). This is, in principle at 
least, the relation between the Fourier modes $\omega_n$ and the corresponding finite diffeomorphism. It is 
the BMS$_3$ analogue of the relation \eqref{rapidity} that we displayed in the Poincar\'e case.\\

The basis of plane waves $|p(\vf),s\ket$ has the virtue of making BMS$_3$ representations manifestly unitary 
\cite{Barnich:2014kra}, since their scalar product takes the form \eqref{norm-waves}. The only difference is 
that now the measure $\mu$ is a path integral measure on a supermomentum orbit; such measures were shown to 
exist in \cite{ShavgulidzeBis}. In addition, this setup provides a simple argument for showing that the 
representation is irreducible. Indeed, the supermomenta $p(\vf)$ of these states span a superrotation orbit; 
the latter is a homogeneous space for the Virasoro group, so any plane wave can be mapped on any other one 
thanks to a suitably chosen superrotation. If we think of plane waves as a basis of the space of the 
representation, then this property of transitivity implies that the space of the representation admits no 
non-trivial invariant subspace, which is to say that the representation is irreducible. We will also expose 
further arguments for irreducibility in sect.~\ref{sec:ultra}.

\subsubsection*{Vacuum module}

The vacuum $\mathfrak{bms}_3$ module can be characterised in a similar way, starting from a 
vacuum state $|0\rangle$ such that
\be
P_m|0\rangle=0\text{ for all }m\in\mathbb{Z}
\quad\text{and}\quad
J_n|0\rangle=0\text{ for }n=-1,0,1.\label{eq:BMSVac}
\ee
Here the condition $P_0|0\rangle=0$ says that the vacuum has zero mass, while the 
additional conditions $J_{\pm1}|0\rangle=0$ reflect Lorentz-invariance. They imply that the stability group 
of the constant momentum $p(\vf) = M - c_2/24$ is enhanced for $M=0$. In the language of induced 
representations, this means that the little group of the vacuum is the Lorentz group instead of the group of 
rotations.\\

If we were dealing with the Poincar\'e algebra, such conditions would produce a trivial representation. Here, 
by contrast, there exist non-trivial ``boosted vacua'' of the form \eqref{eq:BoostedStatesDef}, where now 
$n_1,...,n_N$ are integers different from $-1,0,1$. These vacua are analogous to the boundary gravitons of 
AdS$_3$ \cite{Maloney:2007ud}. The fact that the vacuum is not invariant under the full BMS 
symmetry, but only under its Poincar\'e subgroup, implies that the boosted states \eqref{eq:BoostedStatesDef} 
(with all $n_i$'s $\neq-1,0,1$) can be interpreted as Goldstone-like states associated with broken symmetry 
generators.
Equivalently each state (\ref{eq:BoostedStatesDef}) can be 
interpreted as a vacuum dressed with the three-dimensional analogue of soft graviton degrees of freedom 
created by superrotations.\footnote{Since three-dimensional gravity has no local degrees of freedom, the 
notion of ``soft
graviton'' is ambiguous. (There are no genuine gravitons whose zero-frequency limit would be
soft gravitons.) Our viewpoint here is that ``soft" degrees of freedom coincide with the boundary degrees of 
freedom defined by non-trivial asymptotic symmetries, in accordance with their relation to soft 
theorems \cite{Strominger:2013jfa}.} This also provides a natural interpretation for non-vacuum BMS$_3$ 
representations as particles dressed with soft gravitons. It should be noted that, in contrast to the 
realistic four-dimensional case, supertranslations here do {\it not} create new states.

\subsection{Ultrarelativistic limit of Virasoro modules}\label{sec:ultra}

In analogy with the discussion in sect.\ \ref{Repsl2},  $\mathfrak{bms}_3$ modules emerge as limits of 
irreducible unitary representations of the local conformal algebra, which are built as tensor products of 
irreducible Verma modules of the Virasoro algebra. We still denote the generators of the local conformal 
algebra by two sets of commuting $\cL_m$ and $\bar{\cL}_m$ as in \eqref{so(2,2)}, but now $m \in \ZZ$ and the 
generators obey the centrally extended algebra
\begin{subequations} \label{Virasoro}
\bea
[\cL_m,\cL_n] & = & (m-n)\, \cL_{m+n} + \frac{c}{12}\, m(m^2-1)\d_{m+n,0} \, , \\[1pt]
[\bar{\cL}_m,\bar{\cL}_n] & = & (m-n)\, \bar{\cL}_{m+n}+ \frac{\bar{c}}{12}\, m(m^2-1)\d_{m+n,0} \,.
\eea
\end{subequations}
Highest weight representations of this algebra are built upon an eigenstate 
$|h,\bar{h}\ket$ of $\cL_0$ and $\bar{\cL}_0$ that satisfies
\be \label{highest-weight}
\cL_n | h ,\bar{h} \ket=0\,, \qquad \bar{\cL}_n | h ,\bar{h} \ket=0 \, \quad\text{when }n>0\, .
\ee
The carrier space of the representation is then spanned by the states
\be\label{eq:VirasoroVermaModule}
\cL_{-n_1} \cdots \cL_{-n_k} \bar{\cL}_{-\bar{n}_1} \cdots \bar{\cL}_{-\bar{n}_l} | h , \bar{h} \ket  
\ee
with $n_1 \geq n_2 \geq \cdots \geq n_k > 0$ and a similar ordering for the $\bar{n}_i > 0$.
As suggested by the analysis in sect.~\ref{Repsl2}, we will be interested in large values of $h$ and 
$\bar{h}$, for which these representations are irreducible. In addition the standard hermiticity condition
\be
(\cL_m)^\dagger = \cL_{-m}
\label{cidager}
\ee
yields a scalar product on this space, allowing one to discuss unitarity.\\

As for the Poincar\'e case, one can define the new generators \eqref{lpj} and rewrite this vector space in 
the basis \eqref{eq:BoostedStatesDef}, where $M$ and $s$ are the eigenvalues of $P_0$ and $J_0$ related to $h$ 
and $\bar{h}$ by \eqref{ms}. The change of basis is again invertible because no $J_n$ annihilates the vacuum. 
Each representation of the conformal algebra is still specified by an analogue of \eqref{matrix-elements}, 
where now each state is labelled by the quantum numbers $n_i$, $\bar{n}_j$ and the matrices $\mathsf{P}^{(n)}$ 
and $\mathsf{J}^{(n)}$ also depend on the central charges $c_1$ and $c_2$ defined by
\be \label{ultra-c}
c_1 = c - \bar{c} \, , \qquad c_2 = \frac{c+\bar{c}}{\ell} \, .
\ee
As before, only negative powers of $\ell$ enter $\mathsf{P}^{(n)}$ via the highest-weight conditions 
\eqref{highest-weight} written in the new basis:
\be
\left( P_{\pm n} \pm \frac{1}{\ell} J_{\pm n} \right) | h, \bar{h} \ket = 0 \, .
\ee
A limit $\ell \to \infty$ performed at fixed $M$, $s$ and $c_1$, $c_2$ (rather than e.g.\ at fixed $h$, 
$\bar{h}$) then gives the $\mathfrak{bms}_3$ module that we built from scratch in sect.~\ref{sec:bms3modules}. 
Note, in particular, that the highest-weight state \eqref{highest-weight} is mapped to the rest-frame state 
\eqref{rest-bms} in this limit. In this sense a $\mathfrak{bms}_3$ module is just a high-energy limit of the 
tensor product of two Virasoro modules. Since Virasoro representations are irreducible for large $h$, it is 
reasonable to expect that the same is true of $\mathfrak{bms}_3$ modules.

\subsection{Galilean limit of Virasoro modules}\label{sec:galileo}

In this section we consider another possible group contraction, to be interpreted as a non-relativistic limit 
of the conformal symmetry. This limiting procedure yields an infinite-dimensional extension of the Galilean 
algebra known as Galilean conformal algebra or $\mathfrak{gca}_2$ (see e.g.\ \cite{Bagchi:2009pe}), which is 
isomorphic to $\mathfrak{bms}_3$. In spite of the algebras being isomorphic, the representations that one 
obtains in the ultrarelativistic or Galilean limits are significantly different. In particular, the Galilean 
contraction we are going to review generically leads to non-unitary representations \cite{Bagchi:2009pe}.\\

As in the previous section, the generators $\mathcal{L}_n$ and $\bar{\mathcal{L}}_n$ satisfy the algebra 
\eqref{Virasoro}, and we consider Virasoro highest-weight representations as in \eqref{highest-weight}. In 
order to perform the nonrelativistic limit we introduce a dimensionless contraction parameter $\epsilon$ and 
the new generators
    \begin{equation}\label{eq:GCALinearCombinations}
        M_n\equiv\epsilon\left(\bar{\mathcal{L}}_n-\mathcal{L}_n\right),\qquad
        L_n\equiv\bar{\mathcal{L}}_n+\mathcal{L}_n\,.
    \end{equation}
We stress that the combinations of $\cL_m$'s appearing in this definition are different from those of the 
ultrarelativistic contraction \eqref{lpj}.
In this basis the conformal algebra reads
\begin{subequations} \label{galilean-algebra}
\bea
\,[L_m,L_n] & = & (m-n)\,L_{m+n} +\frac{c_L}{12}\,m(m^2-1)\,\delta_{m+n,0} \,, \\
\,[L_m,M_n] & = & (m-n)\,M_{m+n} +\frac{c_M}{12}\,m(m^2-1)\,\delta_{m+n,0} \,, \\
\,[M_m,M_n] & = & \e^2\left((m-n)\,L_{m+n} +\frac{c_L}{12}\,m(m^2-1)\,\delta_{m+n,0}\right)\,, \label{[M,M]}
\eea
\end{subequations}
where the central charges are given by
\begin{equation} \label{c-galileo}
c_L=\bar{c}+c\,,\qquad c_M=\epsilon\left(\bar{c}-c\right)\,.
\end{equation}
In the limit $\e \to 0$ one obtains an algebra isomorphic to \eqref{jpc}.\\

We denote the eigenvalues of $M_0$ and $L_0$ on a highest-weight state $|h,\bar{h}\rangle$ by
    \begin{equation}
        \Delta=\bar{h}+h\, ,\qquad\xi=\epsilon\left(\bar{h}-h\right) ,
    \end{equation}
and we use them to label the state as $|\Delta,\xi\rangle$. In terms of the operators 
\eqref{eq:GCALinearCombinations} the highest-weight conditions \eqref{highest-weight} become
\begin{equation}\label{eq:GCAHighestWeight}
        L_n|\Delta,\xi\rangle=0\,,\qquad M_n|\Delta,\xi\rangle=0\,,\quad n>0 \, .
    \end{equation}
Note that these constraints hold for any value of $\e$, including the limit $\e \to 0$.
One can then consider the descendant states
\begin{equation}\label{eq:GCAStates}
        |\{\mathfrak{l}_i\},\{\mathfrak{m}_j\}\ket =
        L_{-\mathfrak{l}_1}\ldots L_{-\mathfrak{l}_i}
        M_{-\mathfrak{m}_1}\ldots M_{-\mathfrak{m}_j}|\Delta,\xi\rangle \, ,
    \end{equation}
with $\mathfrak{l}_1\geq\ldots\geq\mathfrak{l}_i > 0$ and $\mathfrak{m}_1\geq\ldots\geq\mathfrak{m}_j > 0$, 
and compute the matrix elements of the operators $M_n$ and $L_n$ in this basis, by using the commutators 
\eqref{galilean-algebra}. Only positive powers of $\e^2$ appear due to \eqref{[M,M]}, while in contrast to the 
ultrarelativistic case the highest-weight conditions \eqref{eq:GCAHighestWeight} do not bring any power of 
$\e$. The matrix elements also depend on the central charges $c_L$ and $c_M$ of \eqref{c-galileo}. In the 
limit $\epsilon\rightarrow0$ at $\Delta$, $\xi$, $c_L$ and $c_M$ fixed one finds the same matrix elements that 
one would obtain by working directly with the $\mathfrak{gca}_2$ algebra.\\

We stress that the highest-weight conditions \eqref{eq:GCAHighestWeight} significantly differ from the 
rest-frame conditions \eqref{highest-weight} that we obtained in the ultrarelativistic limit. Consequently, 
the corresponding representations have very different features. In the Galilean case one can readily define a 
scalar product by imposing the hermiticity conditions $(M_m)^\dagger = M_{-m}$ and $(L_m)^\dagger = L_{-m}$. 
This allows one to compute $\bra \{\mathfrak{l}_i\},\{\mathfrak{m}_j\} 
|\{\mathfrak{l}_k\},\{\mathfrak{m}_l\}\ket$ by taking advantage of \eqref{eq:GCAHighestWeight}. One realises 
in this way that, in contrast with $\mathfrak{bms}_3$ modules, these representations are typically reducible 
and non-unitary \cite{Bagchi:2009pe}.

\section{Higher-spin modules in flat space}\label{sec:hs}

We now turn to the higher-spin analogue of the algebraic constructions described above. For concreteness and 
simplicity we focus on the spin-3 extension of $\mathfrak{bms}_3$ but our considerations apply, mutatis 
mutandis, to other higher-spin extensions as well. We will start by defining the quantum flat $\cW_3$ algebra 
as an ultrarelativistic limit of $\cW_3\oplus\cW_3$, which will produce a specific ordering of operators in 
the non-linear terms of the commutators. Sect.\ \ref{sufawa} will then be devoted to the construction of 
induced modules along the lines described above for Poincar\'e and $\mathfrak{bms}_3$, and we will see there 
that the ordering that emerges in the ultrarelativistic limit may be seen as a normal ordering with respect to 
rest-frame conditions. Along the way we will compare our results to those of the non-relativistic limit 
described in \cite{Grumiller:2014lna}, and we will see that the two limits lead to different quantum algebras.

\subsection{Spin-3 extension of the $\mathfrak{bms}_3$ algebra}

One can add to the $\mathfrak{bms}_3$ algebra two sets of generators $W_n$ and $Q_n$ that transform under 
superrotations as the modes of primary fields of conformal weight 3. This gives a non-linear algebra that can 
be obtained as an \.In\"on\"u-Wigner contraction of the direct sum of two $\cW_3$ algebras. This contraction 
has been discussed at the \emph{semiclassical} level in \cite{Afshar:2013vka,Gonzalez:2013oaa} and a Galilean 
limit of the quantum algebra has been considered in \cite{Grumiller:2014lna}. Here we are interested instead 
in an ultrarelativistic limit of $\cW_3 \oplus \cW_3$. 
The key difference between the Galilean and ultrarelativistic contractions is that the latter mixes 
generators with positive and negative mode numbers, whereas the former does not. For linear algebras, such as 
the Virasoro algebra for example, this does not pose a problem and thus the two contractions yield isomorphic 
algebras. As soon as non-linear algebras are involved in the contraction, however, Galilean and 
ultrarelativistic limits do not necessarily yield isomorphic (quantum) algebras anymore.

\subsubsection*{Ultrarelativistic contraction}

The quantum $\cW_3$ algebra is spanned by two sets of generators $\cL_m$ and $\cW_m$ ($m\in\mathbb{Z}$) whose 
commutation relations read
\begin{subequations}\label{eq:QuantumW3}
\begin{align}
[\mathcal{L}_m,\,\mathcal{L}_n] &= 
(m-n)\,\mathcal{L}_{m+n}+\frac{c}{12}\,(m^3-m)\,\delta_{m+n,\,0}\,, \\
[\mathcal{L}_m,\,\mathcal{W}_n] &= (2m-n)\,\mathcal{W}_{m+n}\,, \\
[\mathcal{W}_m,\,\mathcal{W}_n] &= (m-n)(2m^2+2n^2-mn-8)\,\mathcal{L}_{m+n} + 
\frac{96}{c+\tfrac{22}{5}}\,(m-n)\,\colon\!{\mathcal{L}}{\mathcal{L}}\colon\!_{m+n} \nonumber \\ 
&\quad + \frac{c}{12}\,(m^2-4)(m^3-m)\,\delta_{m+n,\,0}\,,
\end{align}
\end{subequations}
with the usual normal ordering prescription\footnote{The term linear in $\mathcal{L}_m$ ensures that the 
resulting normal-ordered operator $\colon\!{\mathcal{L}}{\mathcal{L}}\colon\!_{m}$ is quasi-primary with 
respect to the action of $\mathcal{L}_m$'s.}
\begin{equation}\label{eq:VirasoroNO}
\colon\!{\mathcal{L}}{\mathcal{L}}\colon\!_{m}
=
\sum_{p\geq -1} \mathcal{L}_{m-p}\mathcal{L}_p 
+
\sum_{p< -1} \mathcal{L}_p{\mathcal{L}}_{m-p}
-
\frac{3}{10} (m+3)(m+2) \mathcal{L}_m\,.
\end{equation}
The standard hermiticity conditions on the generators of this algebra are
\be
(\cW_m)^{\dagger}=\cW_{-m}
\ee
together with \eqref{cidager}; these conditions must hold in any unitary representation of the $\cW_3$ 
algebra.\\

We consider a direct sum $\cW_3\oplus\cW_3$ where the generators and the central charge of the other copy of 
$\mathcal{W}_3$ will be denoted with a bar on top ($\bar{\mathcal{L}}_m,\,\bar{\mathcal{W}}_m$ and $\bar{c}$). 
Introducing a length scale $\ell$ (to be interpreted as the AdS$_3$ radius), we define new generators $P_m$ 
and $J_m$ as in \eqref{lpj} together with  
\be
\label{eq:fshsg12}
W_m\equiv\,\mathcal{W}_m-\bar{\mathcal{W}}_{-m},
\qquad
Q_m\equiv\frac{1}{\ell}\left(\mathcal{W}_m+\bar{\mathcal{W}}_{-m}\right).
\ee
We also define central charges $c_1$ and $c_2$ as in \eqref{ultra-c}. In the limit $\ell \to \infty$, and 
provided the central charges scale in such a way that both $c_1$ and $c_2$ be finite, one finds that $J_m$ and 
$P_m$ satisfy the brackets \eqref{jpc} and
\begin{subequations} \label{comm-W}
\begin{alignat}{5}
[J_m,\, W_n] & = (2m-n) W_{m+n}\,, \qquad &
[J_m,\, Q_n] & = (2m-n) Q_{m+n}\,, \\  
[P_m,\, W_n] & = (2m-n) Q_{m+n}\,, \qquad &
[P_m,\, Q_n] & = 0\,.
\end{alignat}
The remaining brackets involving higher-spin generators are
\begin{align}
[W_m,\, W_n] &= (m-n)(2 m^2 + 2 n^2 - mn -8) J_{m+n} 
		+\frac{96}{c_2}\,  (m-n) \Lambda_{m+n} \nn \\
		& \quad - \frac{96\,c_1 }{c_2^2}\,  (m-n) \Theta_{m+n} 
		+\frac{c_1}{12}\, (m^2-4)(m^3-m)\, \delta_{m+n,\,0}\,, \label{[W,W]} \\
[W_m,\, Q_n] &= (m-n)(2 m^2 + 2 n^2 - mn -8) P_{m+n} 
		+ \frac{96}{c_2}\,  (m-n) \Theta_{m+n} \nonumber \\
		& \quad +\frac{c_2}{12}\, (m^2-4)(m^3-m)\, \delta_{m+n,\,0} \,, \\
[Q_m,\, Q_n] &=	0 \, ,
\end{align}
\end{subequations}
where we have introduced the following notation for non-linear terms:
\begin{equation}\label{eq:FW3NonLinTermsUR}
		\Theta_m  \equiv\sum_{p=-\infty}^\infty P_{m-p} P_p \,,\qquad
		\Lambda_m \equiv\sum_{p=-\infty}^\infty \left( P_{m-p} J_p + J_{m-p} P_p \right) .
\end{equation}
One can check that with this definition the algebra \eqref{jpc}, \eqref{comm-W} satisfies Jacobi identities. 
We will call this algebra the (quantum) \emph{flat $\cW_3$ algebra}. In any unitary representation, its 
generators satisfy the hermiticity conditions
\be
(Q_m)^{\dagger}=Q_{-m}\,,
\quad
(W_m)^{\dagger}=W_{-m}\,.
\ee
supplemented with \eqref{herm} for $m\in\mathbb{Z}$.\\

The expressions \eqref{eq:FW3NonLinTermsUR} for the quadratic terms follow from the identities
\begin{subequations}
\bea
\colon\!{\cL}{\cL}\colon\!_{m} \,+\, \colon\!{\bar{\cL}}{\bar{\cL}}\colon\!_{-m}
& = & \frac{\ell^2}{2}\, \Th_m + \cO(\ell) \, , \\
\colon\!{\cL}{\cL}\colon\!_{m} \,-\, \colon\!{\bar{\cL}}{\bar{\cL}}\colon\!_{-m} 
& = & \frac{\ell}{2}\, \L_m + \cO(1) \, .
\eea
\end{subequations}
Note, in particular, that both the linear term in \eqref{eq:VirasoroNO} and the mixing between positive and 
negative modes in \eqref{lpj}-\eqref{eq:fshsg12} are necessary to reorganize the sum of quadratic terms with 
the precise order  of \eqref{eq:FW3NonLinTermsUR}. We shall see in sect.~\ref{sufawa} that 
\eqref{eq:FW3NonLinTermsUR} can be considered as a normal-ordered polynomial with respect to our definition of 
the vacuum.

\subsubsection*{Galilean contraction}

By contrast, a Galilean contraction of $\cW_3\oplus\cW_3$ can be obtained by defining central charges $c_L$ 
and $c_M$ as in \eqref{c-galileo}, introducing new generators $M_m$ and $L_m$ as in 
\eqref{eq:GCALinearCombinations} and writing
\be
Q_m \equiv \e \left(\bar{\cW}_m - \cW_m \right) , \qquad W_m \equiv \bar{\cW}_m + \cW_m \, .
\ee
In the limit $\e \to 0$ one obtains brackets of the same form as in \eqref{comm-W} after the substitutions 
$J_m \to L_m$, $P_m \to M_m$ and $c_1 \to c_L$, $c_2 \to c_M$. However, there are two important differences: 
the coefficient in front of $\Th_{m+n}$ in \eqref{[W,W]} contains a shifted central charge $c_L + 44/5$ and 
the quadratic terms read
\begin{subequations} \label{noligal}
\bea
\Th_m
& = &
\sum_{p = - \infty}^\infty M_{p} M_{m-p} \, , \\
\L_m
& = &
\sum_{p\geq-1}
\left(M_{m-p}L_p+L_{m-p}M_p\right)
+
\sum_{p<-1}
\left(M_pL_{m-p}+L_pM_{m-p}\right)\nn\\
&   &
-\,
\frac{3}{5}\,(m+3)(m+2)M_m
\eea
\end{subequations}
(see eq.~(2.18) of \cite{Grumiller:2014lna}). They can be interpreted as normal-ordered operators with 
respect to a highest-weight vacuum defined by conditions of the type \eqref{eq:GCAHighestWeight} with 
$\Delta=\xi=0$. These differences show that the two contractions lead to different \emph{quantum} algebras, 
despite the fact that the corresponding classical algebras coincide.\footnote{An interesting problem is to 
understand if these algebras are merely different because of an unfortunate choice of basis, or if they are 
genuinely distinct in the sense that they are not isomorphic. We will not address this issue here.} Thus, in 
the presence of higher-spin fields, the difference between ultrarelativistic and Galilean limits manifests 
itself directly in the symmetry algebras and not only at the level of the representations surviving in the 
limit.\\

In the following we restrict our attention to irreducible unitary representations of the ultrarelativistic 
quantum algebra \eqref{comm-W}, built once again according to the induced module prescription. On the other 
hand, highest-weight representations of Galilean contractions of two copies of non-linear $\mathcal{W}$ 
algebras were discussed in \cite{Grumiller:2014lna}, where it was shown that unitary representations with 
higher-spin states do not exist.

\subsection{Flat $\cW_3$ modules}
\label{sufawa}

The non-linearities in \eqref{comm-W} make standard group-theoretic techniques inapplicable to the 
construction of representations of the flat $\cW_3$ algebra. In spite of this limitation, it was suggested in 
\cite{Campoleoni:2015qrh} that unitary irreducible representations can be built as Hilbert spaces of 
wavefunctionals defined on orbits of extended supermomenta
\be
p(\vf) = 
\sum_{n \in \ZZ} p_n e^{in\vf} \, , \qquad 
q(\vf) = \sum_{n \in \ZZ} q_n e^{in\vf} \, , 
\ee
where $p(\vf)$ is the Bondi mass aspect/supermomentum already introduced in \eqref{supermomentum}, while 
$q(\vf)$ is its spin-3 analogue. The latter is a function on the circle transforming as a cubic density (or 
equivalently as a current of spin 3) under superrotations. In analogy with the $\mathfrak{bms}_3$ case, we 
will further assume that the Hilbert space of wavefunctionals on a given orbit admits a basis of eigenstates 
of $P_m$, $Q_m$, denoted by $|p(\vf),q(\vf)\ket$, where for brevity we omitted the spin labels (such as $s$ in 
\eqref{rest-bms}) that are fixed in any given irreducible representation. This proposal was tested in 
\cite{Campoleoni:2015qrh}, for arbitrary spin, by matching suitable products of one-loop higher-spin partition 
functions with group characters derived using the Frobenius formula that follows from the orbit construction. 
In the remainder of this section we investigate the algebraic counterpart of that proposal.

\subsubsection*{Massive modules}

We focus on orbits that contain a constant extended supermomentum $p(\vf) = M - c_2/24$, $q(\vf) = q_0$. The 
Hilbert space of the corresponding representation then contains a plane wave state $|M,q_0\ket$ that satisfies
\begin{subequations} \label{restW}
\be 
P_m |M,q_0 \ket = 0 \, , \qquad 
Q_m |M,q_0 \ket = 0 \, \quad \mathrm{for}\ m \neq 0 \, ,
\ee
and is an eigenstate of zero-mode charges:
\begin{alignat}{5}
P_0 |M,q_0 \ket & = M |M,q_0 \ket \, , \qquad 
& J_0 |M,q_0 \ket & = s |M,q_0 \ket \, , \\
Q_0 |M,q_0 \ket & = q_0 |M,q_0 \ket \, , \qquad 
& W_0 |M,q_0 \ket & = w |M,q_0 \ket \, .
\end{alignat}
\end{subequations}
Here $M$ and $s$ are the mass and spin labels encountered earlier, while $q_0$ and $w$ are their spin-3 
counterparts. As before we will call $|M,q_0 \ket$ the \emph{rest-frame state} of the representation.\\

The conditions \eqref{restW} define a one-dimensional representation of the subalgebra spanned by 
$\{P_m,Q_m,J_0,W_0\}$. They can be used to define an induced module $\cH_{M,q_0}$ with basis elements
\be \label{Wmodule}
W_{k_1}\cdots W_{k_m}J_{l_1}\cdots J_{l_n}|M,q_0 \ket \, ,
\ee
where $k_1 \geq \cdots \geq k_m$ and $l_1 \geq \cdots \geq l_n$ are non-zero integers. This provides an 
explicit representation of the quantum flat $\cW_3$ algebra. Note that the presence of non-linearities in the 
commutators \eqref{comm-W} does not affect the construction of the induced module, which involves the 
universal enveloping algebra anyway.\\

As in the previous examples, the basis \eqref{Wmodule} is very useful to prove the existence of a given 
representation, but not very illuminating if one wants to understand its properties. Following our mantra, we 
thus move to a basis of eigenstates of supermomentum by acting on the rest-frame state as
\be \label{UflatW}
|p(\vf),q(\vf)\ket=U(\o,\O)|M,q_0 \ket \, ,
\ee
where
\be
U(\o,\O)=\exp\left(i\sum_{n\in\ZZ^{*}}( \o_{n} J_{n} +\O_{n} W_{n}) \right)
\quad
\text{with }\omega_n^*=\omega_{-n},\;\Omega_n^*=\Omega_{-n}
\ee
is a unitary operator implementing a finite higher-spin superrotation. The complex coefficients $\omega_n$ 
and $\Omega_n$ can be interpreted as the Fourier modes of the tensor fields $\o(\vf)\partial_\vf$ and 
$\O(\vf)\partial_\vf \partial_\vf$ on the circle.\\

To prove that the states \eqref{UflatW} diagonalise all $P_m$'s and $Q_m$'s we can use the 
Baker-Campbell-Hausdorff formula as in \eqref{aqui} to obtain
\begin{subequations} \label{nested}
\bea
P_m |p(\vf),q(\vf) \ket & = & U\cdot \exp\left( i\,\textrm{ad}_{\o,\O}\right) [P_m] |M,q_0\ket \, ,\\
Q_n |p(\vf),q(\vf) \ket & = & U\cdot \exp\left( i\,\textrm{ad}_{\o,\O}\right) [Q_n] |M,q_0\ket \,,
\eea
\end{subequations}
where
\be
\textrm{ad}_{\o,\O}[\,\cdot\,] = \sum_{n\in\ZZ^*} [\, \o_{n} J_{n} +\O_{n} W_{n} , \,\cdot \,] \, .
\ee
Inspection of \eqref{comm-W} then shows that the nested commutators that appear in \eqref{nested} always 
produce either $P_n$ or $Q_n$ modes. As a result the right-hand sides of \eqref{nested} turn into series of 
products of $P_n$ and $Q_n$ operators, which act diagonally on the rest-frame state by virtue of the 
rest-frame conditions \eqref{restW}. Computing the eigenvalue associated with a given set of modes $\o_n$ and 
$\O_n$ requires instead a control of the finite action of $\cW_3$ superrotations, that was discussed e.g.\ in 
\cite{Gomis:1994rz}.\\

As before the plane waves \eqref{UflatW} provide an orthonormal basis of the Hilbert space of the 
representation, thus making unitarity manifest (provided a suitable measure exists on the orbit). 
Irreducibility can be inferred from the same argument we used for $\mathfrak{bms}_3$: by construction, a 
supermomentum orbit is a homogeneous space for the action of superrotations, and this carries over to the 
higher-spin setting. This implies that $\cW_3$ superrotations can map any plane wave state on any other one, 
which in turn implies that the space of the representation has no non-trivial invariant subspace.

\subsubsection*{Vacuum module}

The vacuum module of the flat $\cW_3$ algebra can be built in direct analogy to its $\mathfrak{bms}_3$ 
counterpart discussed around \eqref{eq:BMSVac}. The only subtlety is the enhancement of the little group, 
which leads to additional conditions on superrotations. Indeed the vacuum state $|0\ket$ is now an eigenstate 
of all modes $P_m$ and $Q_m$ with zero eigenvalue, and satisfies in addition
\be
J_n|0\ket=0\;\text{ for }n=-1,0,1,
\quad
W_m|0\ket=0\;\text{ for }m=-2,-1,0,1,2.
\ee
These conditions ensure that the vacuum is invariant under the $\mathfrak{sl}(3,\RR)$ wedge algebra of the 
$\cW_3$ subalgebra (which includes in particular the Lorentz algebra). The corresponding module can then be 
built as usual by acting with higher-spin superrotation generators on the vacuum state and producing states of 
the form \eqref{Wmodule}, where now all $l_i$'s must be different from $-1,0,1$ and all $k_i$'s must be 
different from $-2,-1,0,1,2$.\\

The definition of the flat $\cW_3$ vacuum allows us to interpret the quadratic terms in 
\eqref{eq:FW3NonLinTermsUR} as being normal-ordered. Indeed, the expectation value of the operators $\Theta_n$ 
and $\Lambda_n$ vanishes on the vacuum $|0\ket$:
    \begin{equation}\label{eq:RestFrameNO}
        \langle 0 |\Theta_n|0 \rangle=\langle 0|\Lambda_n|0\rangle=0.
    \end{equation}
By contrast, for a highest-weight vacuum of the type \eqref{eq:GCAHighestWeight} one obtains non-vanishing 
expectation values. Thus the additional non-linear structure introduced by higher spins stresses once more 
that the natural representations to be considered in the ultrarelativistic limit are the induced ones 
discussed above, rather than the highest-weight ones of \cite{Bagchi:2009pe,Grumiller:2014lna}.\\
    
These considerations appear to be a robust feature of ``flat $\cW$ algebras''. Ultrarelativistic contractions 
of $\cW_N \oplus \cW_N$ algebras always take the form
\be
\text{``flat $\cW_N$''}
=
\cW_N\inplus_{\text{ad}}(\cW_N)_{\text{Ab}}
\ee
and therefore contain an Abelian ideal. In addition the structure constants of the non-linear terms are 
always proportional to inverse powers of the central charge. Indeed, for a non-linear operator of 
$n^\textnormal{th}$ order the structure constants for large $c$ are proportional to $\frac{1}{c^{n-1}}$. When 
expanding them in powers of the contraction parameter $\ell$, this implies that the leading term is 
proportional to $\ell^{1-n}$ thanks to \eqref{ultra-c}. In order to obtain a finite result, it is thus 
necessary that the resulting non-linear operator consists of at least $n-1$ Abelian generators. Terms of this 
kind always have a vanishing expectation value on our rest-frame vacuum, although the precise ordering in the 
polynomial should be fixed by other means, e.g.\ by defining the algebra via a contraction of the quantum 
algebra or by imposing Jacobi identities.

\section{Conclusion}

In this work we have seen how induced representations of groups can be recovered from induced modules of Lie 
algebras. When applied to $\mathfrak{bms}_3$ and its higher-spin extensions, this approach confirms and 
expands the prescription previously described in \cite{Campoleoni:2015qrh}. Each such module consists of a 
rest-frame state and of its ``descendants'' obtained by acting with all superrotation generators. It provides 
an explicit irreducible unitary representation of the corresponding symmetry algebra. It can also be seen as a 
high-energy, high central charge limit of highest weight representations of direct sums of Virasoro or $\cW$ 
algebras.\\

This approach has also allowed us to discuss the differences between ultrarelativistic and Galilean limits of 
Virasoro/$\cW$ representations. As argued above, the ultrarelativistic limit always produces unitary induced 
modules, while the Galilean limit generically produces non-unitary highest-weight representations. In fact, 
higher-spin considerations allow us to see this dichotomy already at the level of Lie algebras, regardless of 
representation theory: the ultrarelativistic limit of the quantum $\cW_3\oplus\cW_3$ algebra produces 
non-linear terms of the type \eqref{eq:FW3NonLinTermsUR}, which are normal-ordered with respect to the 
rest-frame conditions adapted to induced modules. By contrast, its Galilean limit leads to terms such as 
\eqref{noligal}, which are \emph{not} normal-ordered with respect to rest-frame conditions, but are indeed 
normal-ordered with respect to the highest-weight conditions of \cite{Grumiller:2014lna}. We stress that this 
difference is a genuine quantum higher-spin effect: it is not apparent at the classical level, and it does not 
appear in pure gravity either.\\

The difference between ultrarelativistic and Galilean limits also emphasizes their physical distinction: the 
ultrarelativistic limit is adapted to gravity, and more generally to models of fundamental interactions, where 
unitarity is a crucial requirement. In particular, flat space holography (at least in the framework of 
Einstein gravity) should rely on the unitary construction of \cite{Barnich:2014kra,Barnich:2015uva}, as 
confirmed in \cite{Barnich:2015mui,Campoleoni:2015qrh} by the matching between group characters and one-loop 
partition functions. By contrast, the Galilean viewpoint is suited to condensed matter applications and, more 
generally, to situations where unitarity need not hold --- as was indeed argued in \cite{Bagchi:2009pe}.

\section*{Acknowledgements}

The authors are grateful to G.\ Barnich for illuminating discussions on BMS representations. A.C.\ would like 
to thank the Nanyang Technological University in Singapore for hospitality, and the participants in the NTU 
Workshop on Higher-Spin Gauge Theories for feedback on a presentation of some of the results contained in this 
paper. The work of A.C.\ and H.A.G.\ is partially supported by the ERC Advanced Grant ``SyDuGraM'', by 
FNRS-Belgium (convention FRFC PDR T.1025.14 and convention IISN 4.4503.15) and by the ``Communaut\'e 
Fran\c{c}aise de Belgique" through the ARC program. The work of B.O.\ is supported by a doctoral fellowship of 
the Wiener-Anspach Foundation and by the Fund for Scientific Research-FNRS under grant number FC-95570. This 
research of M.R.\ is supported by the FWF projects P27182-N27 and the START project Y435-N16. M.R.\ is also 
supported by a \emph{DOC} fellowship of the Austrian Academy of Sciences, the \emph{Doktoratskolleg Particles 
and Interactions} (FWF project DKW1252-N27) and the FWF project I 1030-N27.

\begin{appendix}

\section{Superrotations and supermomenta}
\label{sesusu}

In this appendix we revisit the arguments surrounding \eqref{coadjoint} according to which supermomenta 
transform under superrotations as Virasoro coadjoint vectors. We also display the exact relation between a 
vector field $\omega(\vf)\partial_{\vf}$ representing an infinitesimal superrotation, and its exponential.

\subsection{Action of superrotations on supermomenta}

We consider a state $|p(\vf),s\ket$ and act on it with a superrotation $f$ using a unitary operator $U[f]$. 
(The discussion below \eqref{eq:Superwaves} is recovered upon taking $|p(\vf),s\ket=|M,s\ket$ and 
$f=\exp[\omega]$, but our argument will be independent of these details.) We wish to show that the state 
$U[f]|p(\vf),s\ket$ is an eigenstate of supermomentum whose eigenvalues are the Fourier modes of the 
supermomentum $f\cdot p$, where the dot denotes the coadjoint representation \eqref{coadjoint}.\\

We start by combining the generators $P_m$ in a ``quantum supermomentum"
\be
P(\vf)\equiv\sum_{m\in\mathbb{Z}}P_me^{im\vf}
\ee
and act with it on the state $U[f]|p(\vf),s\ket$. We find
\be
P(\vf)U[f]|p(\vf),s\ket
=
U[f]\left(U[f]^{-1}P(\vf)U[f]\right)|p(\vf),s\ket\,.
\label{puki}
\ee
Now recall how the supermomentum generators $P_m$ are defined: a supertranslation $\alpha$ is represented by 
a unitary operator
\be
U[\alpha]
=
\exp
\left[
\frac{i}{2\pi}\int_0^{2\pi}d\vf\,\alpha(\vf)P(\vf)
\right]
\label{unitra}
\ee
so we have
\be
P(\vf)
=
-2\pi i\,\frac{\delta}{\delta\alpha(\vf)}
\left.
\left(
U[\alpha]
\right)
\right|_{\alpha=0}
\ee
and the definition of $P_m$ follows after taking the $m^{\text{th}}$ Fourier mode of this expression. Now, 
plugging this definition in \eqref{puki} we get
\bea
P(\vf)U[f]|p(\vf),s\ket
& = &
-2\pi i\,\frac{\delta}{\delta\alpha(\vf)}
\left.
U[f]\left(
U[f]^{-1}
U[\alpha]
U[f]
\right)
\right|_{\alpha=0}
|p(\vf),s\ket\nn\\
\label{wakawaka}
& = &
-2\pi i\,\frac{\delta}{\delta\alpha(\vf)}
\left.
U[f]\left(
U[\sigma_{f^{-1}}\alpha]
\right)
\right|_{\alpha=0}
|p(\vf),s\ket
\eea
where $\sigma$ denotes the adjoint representation of the Virasoro group. In writing this we have used the 
group operation of BMS$_3$,
\be
(f,\alpha)\cdot (g,\beta)
=
(f\circ g,\alpha+\sigma_f\beta)\,.
\ee
Now using \eqref{unitra} we find
\be
U[\sigma_{f^{-1}}\alpha]
=
\exp
\left[
\frac{i}{2\pi}\int_0^{2\pi}d\vf\,(\sigma_{f^{-1}}\alpha)(\vf)P(\vf)
\right]
=
\exp
\left[
\frac{i}{2\pi}\int_0^{2\pi}d\vf\,\alpha(\vf)(\sigma^*_fP)(\vf)
\right]
\ee
where $\sigma^*$ is the dual representation corresponding to $\sigma$
and thus coincides with the coadjoint representation of the Virasoro group. Plugging this in \eqref{wakawaka} 
and taking the functional derivative we finally find
\be
P(\vf)U[f]|p(\vf),s\ket
=
U[f]
(\sigma^*_fP)(\vf)
|p(\vf),s\ket\,.
\ee
Since $|p(\vf),s\ket$ is an eigenstate of supermomentum, it follows that
\be
P(\vf)U[f]|p(\vf),s\ket
=
(\sigma^*_fp)(\vf)
U[f]
|p(\vf),s\ket
\ee
which was to be proven.

\subsection{Exponential superrotations}

As mentioned in footnote \ref{footex}, the exponential map of the Virasoro group is not locally surjective. 
Nevertheless one can consider superrotations given by exponentials of vector fields as in \eqref{expu}. It is 
then natural to ask what is the explicit form of the diffeomorphism $f=\exp[\omega]$ generated by a vector 
field $\omega=\omega(\vf)\partial_{\vf}$.\\

As mentioned below \eqref{coadjoint} the relation between $f$ and $\omega$ is given by the flow of the 
latter. Explicitly, $\omega$ defines a family of integral curves, which are paths $\vf(t)$ on the circle that 
satisfy the differential equation
\be
\frac{d\vf(t)}{dt}=\omega(\vf(t)).
\label{flow}
\ee
Given an initial condition $\vf(0)=\vf_0$, the solution of this equation is unique; this defines a 
one-parameter family of diffeomorphisms $f_t$ such that
\be
f_t(\vf_0)=\vf(t)
\ee
and known as the flow of $\omega$.
In particular, by definition of the exponential map, the diffeomorphism $f=\exp[\omega]$ generated by 
$\omega$ is just $f=f_1$, that is, the flow evaluated at time $t=1$. Using the evolution equation \eqref{flow} 
this can also be restated as the condition
\be
\int_{\vf_0}^{f(\vf_0)}
\frac{d\vf}{\omega(\vf)}
=
1,
\label{expif}
\ee
required to hold for any $\vf_0$ on the circle. This is the exact correspondence between a superrotation 
generator $\omega$ and its exponential $f=\exp[\omega]$. Note that \eqref{expif} automatically ensures 
$f(\vf_0+2\pi)=f(\vf_0)+2\pi$, in accordance with the fact that $f$ is a diffeomorphism of the circle.

\end{appendix}

\end{document}